\documentclass[journal=jacsat,manuscript=article]{achemso}

\usepackage{graphicx}
\usepackage{dcolumn}
\usepackage{bm}
\usepackage{amssymb}
\usepackage{amsmath}
\usepackage{doi}
\usepackage{hyperref}
\usepackage{textcomp}
\usepackage[caption=false]{subfig}

\usepackage{color}
\usepackage[normalem]{ulem}

\title{Cooperative effect of electrons spin polarization in a hybrid nanostructure of a magnetic thin film with adsorbed chiral  molecules studied with non-spin-polarized scanning tunneling microscopy}

\author{Nguyen T. N. Ha}
\affiliation{Solid Surface Analysis, Institute of  Physics, Chemnitz University of Technology, Reichenhainer Str. 70, 09126 Chemnitz. Germany}
\author{L.~Rasabathina}
\affiliation{Functional Magnetic Materials, Institute of Physics, Chemnitz University of Technology, Reichenhainer Str. 70, 09126 Chemnitz, Germany}
\author{O.~Hellwig}
\affiliation{Functional Magnetic Materials, Institute of Physics, Chemnitz University of Technology, Reichenhainer Str. 70, 09126 Chemnitz, Germany}
\alsoaffiliation{ Institute of Ion Beam Physics and Materials Research, Helmholtz-ZentrumDresden-Rossendorf, Bautzner Landstrasse 400, 01328 Dresden, Germany}
\author{A.~Sharma}
\affiliation{Semiconductor Physics, Institute of Physics, Chemnitz University of Technology, Reichenhainer Str. 70, 09126 Chemnitz, Germany}
\author{G.~Salvan}
\affiliation{Semiconductor Physics, Institute of Physics, Chemnitz University of Technology, Reichenhainer Str. 70, 09126 Chemnitz, Germany}
\author{S.~Yochelis}
\affiliation{Department of Applied Physics, Hebrew University of Jerusalem, Israel}
\alsoaffiliation{Center for Nanoscience and Nanotechnology, Hebrew University of Jerusalem, Israel}
\author{Y.~Paltiel}
\affiliation{Department of Applied Physics, Hebrew University of Jerusalem, Israel}
\alsoaffiliation{Center for Nanoscience and Nanotechnology, Hebrew University of Jerusalem, Israel}
\author{L.T.~Baczewski}
\affiliation{Institute of Physics, Polish Academy of Sciences, Al. Lotnikow 32/46, 02-668 Warszawa, Poland}
\author{C.~Tegenkamp}\email{christoph.tegenkamp@physik.tu-chemnitz.de}
\affiliation{Solid Surface Analysis, Institute of  Physics, Chemnitz University of Technology, Reichenhainer Str. 70, 09126 Chemnitz. Germany}

\begin{document}
\today
\clearpage
\begin{abstract}
Polyalanine molecules (PA) with an $\alpha$-helix conformation  gathered recently a lot of interest as the propagation of electrons through the chiral backbone structure comes along with spin polarization of the transmitted electrons. 
By means of scanning tunneling microscopy and spectroscopy at ambient conditions,  PA molecules adsorbed on surfaces of epitaxial magnetic Al$_2$O$_3$/Pt/Au/Co/Au nanostructures with perpendicular anisotropy were studied.  Thereby, a correlation between the PA molecules ordering at the surface with the electron tunneling across this hybrid system as a function of the substrate magnetization orientation as well as the coverage density and helicity of the was observed. 
The highest spin polarization values, P, were found for well-ordered self-assembled monolayers and with a defined chemical coupling of the molecules to the magnetic substrate surface, showing that the current induced spin selectivity is  a cooperative effect.  Thereby, P deduced from the electron transmission along unoccupied molecular orbitals of the helical molecules is larger as compared to values derived from the occupied molecular orbitals. Apparently,  the larger orbital overlap is resulting in a higher electron mobility yielding a higher P value.  By switching the magnetization direction of the Co-layer, it was demonstrated that  the non-spin-polarized  STM  can be used to study chiral molecules with a sub-molecular resolution,  to detect properties of buried magnetic layers and to detect  the spin polarization of the molecules from the change of the magnetoresistance of such hybrid structures.

\end{abstract}

\maketitle

\section{Introduction}

Since the discovery of the so-called chiral induced spin selectivity (CISS) effect, chiral or more precisely helical molecules  attracted much attention in view of molecular-based spin filter materials \cite{Xie2011, Naaman2019, Naaman2020, Waldeck2021}.
Moreover, helical molecules adsorbed on a magnetic thin film are able to induce magnetization reorientation without applying magnetic or electric fields and even show the enantiospecific adsorption depending on the substrate magnetization direction \cite{Nir2020, Meirzada2021, Gosh2018, Tassinari2019}. The spin-polarization found in helical polyalanine molecules (PA) is robust, i.e. in the order of 60-80~\% at ambient conditions \cite{Ghosh2020}, therefore promising for future hybrid spin-devices \cite{Liu2019, NaamanWY2019}. 

The coupling and ordering of the molecules to the magnetic substrate are important in hybrid nanostructures. 
Structural  details  about the density and orientation of the adsorbed molecules are necessary for a comprehensive understanding of the CISS effect. Even for rather complex molecules, like polyalanines, submolecular resolution and information of the morphology on the atomic scale was demonstrated by  scanning tunneling microscopy \cite{Ha2019,HaAu2020,HaDL2020}. 

In order to measure the spin-polarization in helical molecules dc transport experiments down to the scale of single molecules are mandatory  \cite{Xie2011, Aragones2017, Slawig2020}. Moreover, combinations of magnetic contacts with defined magnetization directions are required for quantification of the spin polarization \cite{Aragones2017}. So far, atomic force microscopy (AFM) was used to probe the structure for understanding structure-property relations on the microscopic scale \cite{Kiran2016, Ghosh2020, Ziv2019}. Indeed, using conductive AFM spectroscopy, the current-voltage (I-V) curves provide higher transmission if the helicity direction of the molecules and, concomitantly, spin polarization is matching the direction of the magnetization. However, the spatial resolution of the AFM is usually insufficient to correlate these spectroscopic findings with details of the molecular structure.
 
STM has proven being a powerful  tool to characterize chiral surfaces, visualize chiral molecular structures and provide information of their magnetic and electronic properties at the nanoscale \cite{Wortmann2001, HaDL2020, Ha2019, HaAu2020}.  
Thereby,  spin-polarized STM techniques are often used to determine details of the spin-state and magnetic structure. The STM probes, where a few atoms at a well defined apex define the spin resolution,  are difficult  to make, require ultra-high vacuum conditions and were so far not shown to reveal magnetic contrast under ambient conditions. An alternative is to use ferromagnetic substrates without the need of magnetized tips \cite{Nir2020, BenDor2017, Carmeli2002}. However, such surfaces are prone to oxidation, which will alter the adsorption behavior and ordering of the molecules.

A promising way out of this dilemma is to use Au/Co/Au/chiral-molecules heterostructures, where the spin-state of the electrons within helical molecules adsorbed at the surface of a magnetic substrate can be quantified by means of the  magneto-resistance (MR) \cite{Mondal2021, Nir2020, BenDor2017, Carmeli2002}.  The chemically inert Au-capping layer on top of the magnetic layer allows the use of established chemical concepts, e.g. the thiol-mediated chemisorption of molecules and,  secondly, non-spin-polarized tips can be used to probe the spin-polarization. Usually, high resolution STM experiments are performed on atomically flat surfaces, where the morphology can be  subsequently further improved by various techniques under vacuum conditions. Using the surfaces of Au/Co/Au  nanostructures is challenging, since the growth of these structures come inevitably along with roughening. In order to protect the magnetic properties, the  thermal annealing range is strongly limited to avoid interdiffusion at the interfaces. 

In this study we present  STM and STS measurements on self-assembled PA molecules  adsorbed on the surfaces of Au/Co/Au nanostructures, where we were able to control the molecular structure and determine the spin-polarization via current-voltage (feedback open) and tip height measurements (feedback closed) and compare these with PA layers adsorbed on surfaces of Au-films on mica. Versus former approaches, the high resolution capability of our setup allows 
the correlation of structural properties and molecular orbital selectivity with spin-polarization of the itinerant electrons.  
Moreover, the MR, present at the buried interface, can be elucidated by a change of the resistance of the tunneling junction, present at the surface. Therefore, our setup might be useful to image bulk magnetic states even without the need of spin-polarized STM tips.

\section{Experimental details}

\begin{figure}[tb]
	\includegraphics[width=0.6\columnwidth]{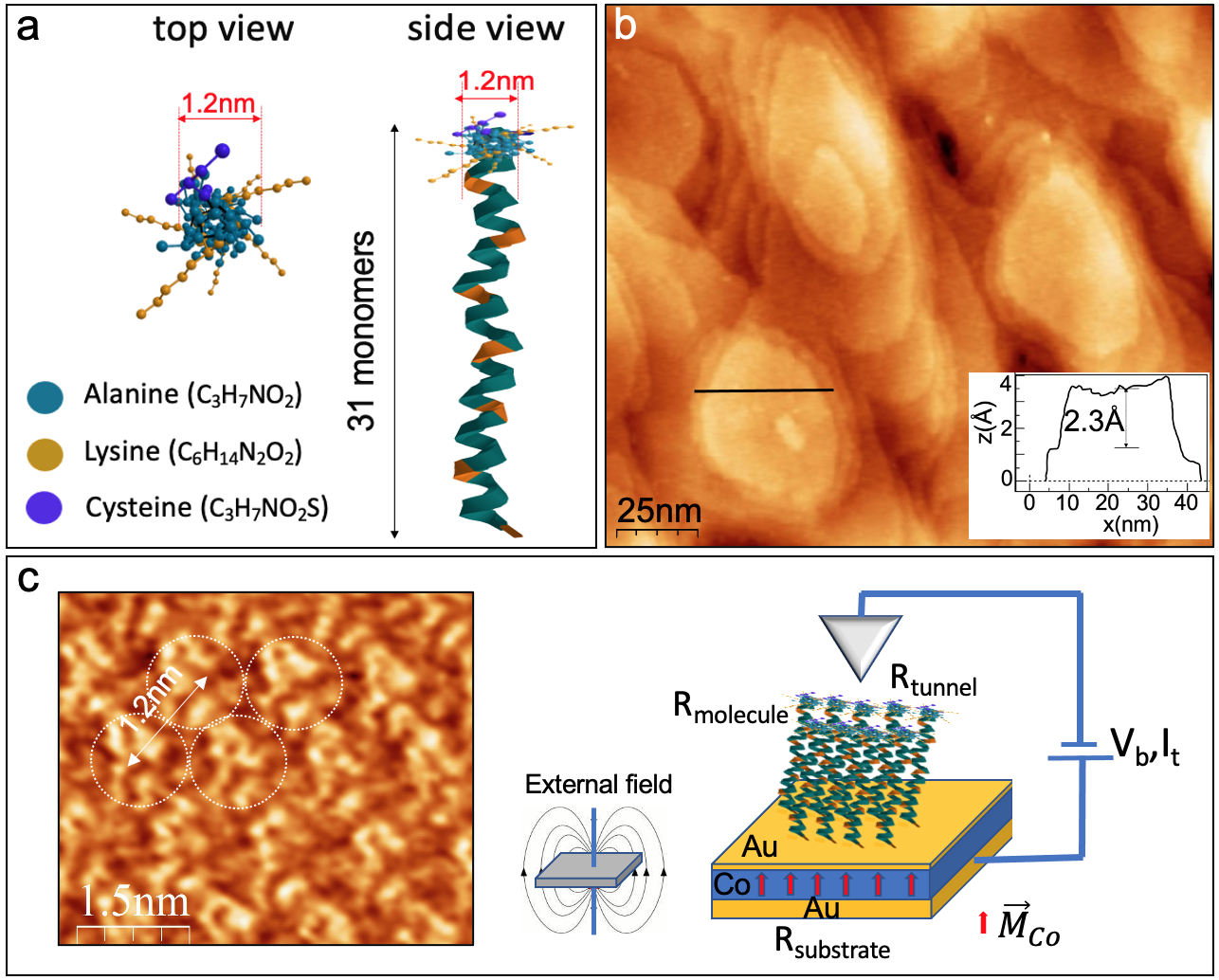}
	\caption{\label{FIG1} a) Model of a helical 31-mer polyalanine  molecule. b) STM image ($80\times 80~nm^2$) of the clean Au/Co/Au surface. Inset: line scan showing a step height of 2.3~\AA~ and flat Au terraces. c) Schematic of our STM setup showing the hybrid system, consisting of a tunneling gap, the helical molecule and the magnetic substrate. The out-of-plane magnetization of the Co-film, $\vec{M}_{Co}$, is switched by an external magnet.  Inset: high resolution  STM image of interdigitated (36-mer) L-PA molecules on  Au(111) films grown on mica \cite{HaAu2020}. The average intermolecular distance of this closed-packed phase is 1.2~nm.}
\end{figure}

The $\alpha$-L and $\alpha$-D polypeptide molecules (L-PA, D-PA) used in this study (commercially available from Aldrich)  comprise the following sequence C[AAAAK]$_6$, where  C, A, and K represent cysteine (C), alanine (A), and lysine (K), respectively. A schematic of a 31-mer PA molecule in shown in Fig.~\ref{FIG1}a).   The magnetic substrate for the STM experiments was a molecular beam epitaxy (MBE) grown epitaxial nanostructure composed of subsequently deposited layers of Au and Co on Pt/sapphire samples, Al$_2$O$_3$/Pt/Au(200~nm)/Co(1.2~nm)/Au(5~nm) (Au/Co/Au in the following), were the Au(111) cap layer of 5~nm thickness prevents the Co layer against  oxidation and provides a defined chemical template for the chemisorption of the cysteine-terminated PA molecules. The Co-layer of 1.2~nm  thickness reveals an out-of-plane magnetization.  
The growth rate of individual layers was not exceeding 0.1 \AA/s. The Pt film was deposited using electron beam evaporation and for the deposition of Au and Co layers effusion cells were used. After deposition of each layer reflection high-energy electron diffraction (RHEED) images were taken to check the epitaxial relations of grown films showing characteristic sharp streaks confirming a 2D growth mode. To assure a good quality of the layers surface the Pt buffer was deposited at 700$^{\circ}$C and the next Au layer was annealed at 200$^{\circ}$C for 2 hours. The Co layer and Au capping layer were finally deposited at room temperature to avoid interdiffusion at the interfaces. The optimized growth procedure resulted in  20-40~nm wide Au(111) flat terraces which are very suitable for an ordered PA molecules adsorption.  A  characteristic  STM image is shown in Fig.~\ref{FIG1}b). Simultaneously the perpendicular anisotropy of the Co layer was maintained. 

In this study, 0.3~mM and 1.0~mM solutions of both L-PA  and D-PA molecules with ethanol as a solvent were used. Self-assembled structures, i.e. monolayers and molecular clusters,  of both enantiomeric PA species were prepared by either drop-casting of PA solution on the substrates  or dipping the substrate into the  PA solution. In the latter case, the sample was removed from the solution and rinsed with absolute ethanol, and subsequently dried with nitrogen gas. The  STM investigations were performed immediately after preparation in order to minimize any degradation effects. 

All scanning tunneling microscopy (STM) and spectroscopy (STS) experiments (Fa. RHK) were performed at  ambient conditions at 300~K using mechanically made (non-spin-polarized) PtIr tips. The dI/dV-V spectra were recorded with a lock-in amplifier (24~meV, 1600~Hz), along with I-V curves, at various locations across the sample and with different  PtIr tips. 

Fig.~\ref{FIG1}c) sketches the main idea for measuring and quantifying the current-induced spin effects. Thereby, the out-of-plane direction of the magnetization of the Co-film, $\vec{M}_{Co}$, was aligned with an external magnet, where a switching was achieved with 40~mT, i.e. by approaching the external magnet to within 2~cm of the sample. 

Both the orientation and strength of the external magnet were calibrated before by means 
of the polar magneto-optical Kerr effect (P-MOKE) spectra of the Au/Co/Au samples recorded for the spectral range of 2.0 eV and 4.5 eV using a home-built MOKE spectrometer. At first, the sample was saturated in the direction normal to the sample surface so the net magnetization vector $\vec{M}_{Co}$ is pointing out of the sample plane by putting it  in close vicinity ($\approx$2 cm) to a permanent magnet. Afterwards, the permanent magnet was removed leaving the sample in a remanent magnetization state $\vec{M}_{Co}=\uparrow$ and the P-MOKE spectrum was recorded. In a similar way the P-MOKE spectrum of the same sample with opposite remanence magnetization direction  ($\vec{M}_{Co}=\downarrow$) was measured. The field strength of ~340 mT was measured at the surface of the permanent magnet using a Gaussmeter. Additionally, we measured the magnetic field strength as a function of distance from the permanent magnet  and found that up to 3 cm from the magnet the field strength was still sufficient to saturate the sample.

\section{Results and discussion}

\subsection{Adsorption of helical PA on magnetic Au/Co/Au substrates: the role of Co-film magnetization orientation and PA molecules concentration}
\begin{figure}[tb]
	\includegraphics[width=0.8\columnwidth]{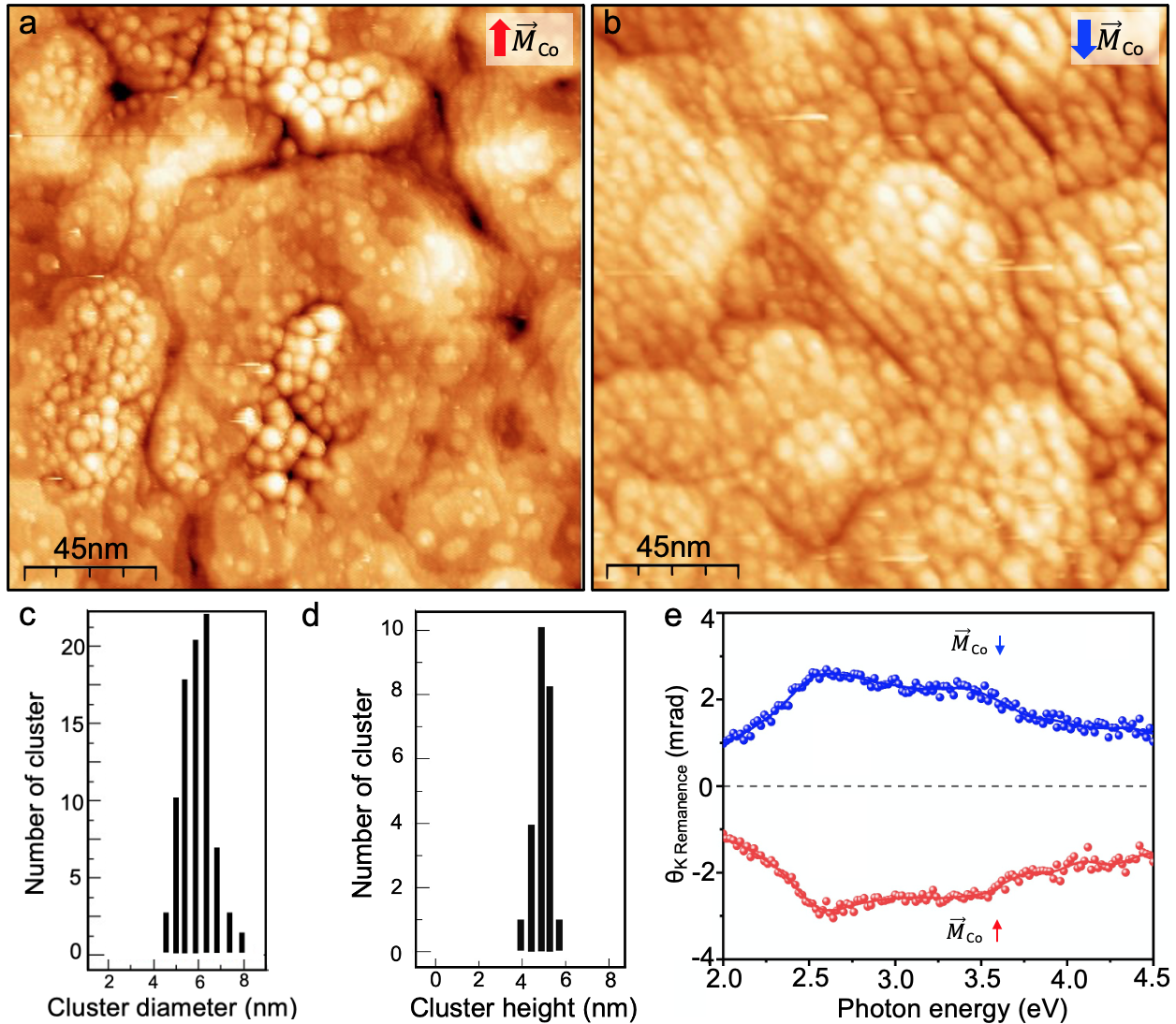}
	\caption{\label{FIG2} Self-assembled structure of L-PA (1~mM) on Au/Co/Au with $\vec{M}_{Co}=\uparrow$ a) and  $\vec{M}_{Co}=\downarrow$  b). The tunneling conditions for both images were  V$_b$=+1.6~V, I$_t$ = 0.26~nA (300~K). c) Cluster diameter distribution deduced from b). d) Cluster height distribution obtained from individual clusters, e.g. visible in panel a). e) The P-MOKE spectra for the PA free Au/Co/Au system of the $\vec{M}_{Co}=\downarrow$ and $\vec{M}_{Co}=\uparrow$ magnetization states are shown by blue and red filled circles, respectively. }
\end{figure}

It was reported that,  depending on the magnetization direction of the substrate, one enantiomer is preferably adsorbed over the other due to the magnetic-exchange interaction of the spin-polarized electrons within the helical molecules with the spin-polarized substrate \cite{Gosh2018}.
In the experiment, shown in Fig.~\ref{FIG2},  we have used this magnetic-exchange interaction in order to probe qualitatively the enantioselective adsorption. As mentioned above, the magnetization orientation of the Co-film, $\vec{M}_{Co}$, was set with the help of an external magnet. The successful magnetization switching of the sample when rotating the permanent magnet was confirmed using the P-MOKE spectrometry technique. Fig.~\ref{FIG2}e)  shows the P-MOKE spectra of the two remanent magnetization states ( $\vec{M}_{Co}=\uparrow$  and $\vec{M}_{Co}=\downarrow$). As expected the corresponding spectra  are mirror images to each other demonstrating that the magnetic field from the permanent magnet is sufficient for switching the magnetization of the buried cobalt layer \cite{Erskine1973}. The spectral line shape and spectral feature are consistent with our previous studies \cite{Sharma2020}.

L-PA were adsorbed from a 1~mM solution on differently magnetized Au/Co/Au samples, shown in Fig.~\ref{FIG2}.
Compared to the PA free  Au/Co/Au substrate (Fig.~\ref{FIG1}b), the liquid-solid interaction results in appearance of clusters  of nearly identical diameter  (6-8~nm, cf. cluster size distribution in Fig.~\ref{FIG2}c). However, the cluster densities for the oppositely magnetized surfaces strongly differ, although the PA concentration as well as the interaction time (2~min.) were the same and the roughness parameters of the Au/Co/Au samples are comparable. Thus the observed differences must originate from the differently magnetized buried Co layers.
While in the case of a downward pointing magnetization  ($\vec{M}_{Co}=\downarrow$) a homogeneous cluster phase with a density of around 0.02/nm$^2$ was obtained (Fig.~\ref{FIG2}b), in the case of  $\vec{M}_{Co}=\uparrow$ the overall coverage is around 3 times lower and less dense so that the flat terrace topography of the PA free  Au/Co/Au surface is still visible.  
As it is obvious from Fig.~\ref{FIG2}a), the clusters are non-homogeneously distributed, thus the magnetic exchange interaction may vary across the surface, e.g. by different heights of the Au-film. However,  defect-mediated adsorption of clusters can also affect the clustering beside the magnetic interaction effect. 
Nevertheless, we will need to confirm  that the clusters are formed only from L-PA molecules and to exclude  parasitic effects like molecules fragmentation or destruction of the $\alpha$-helix structure.


\begin{figure}[t]
	\includegraphics[width=.8\columnwidth]{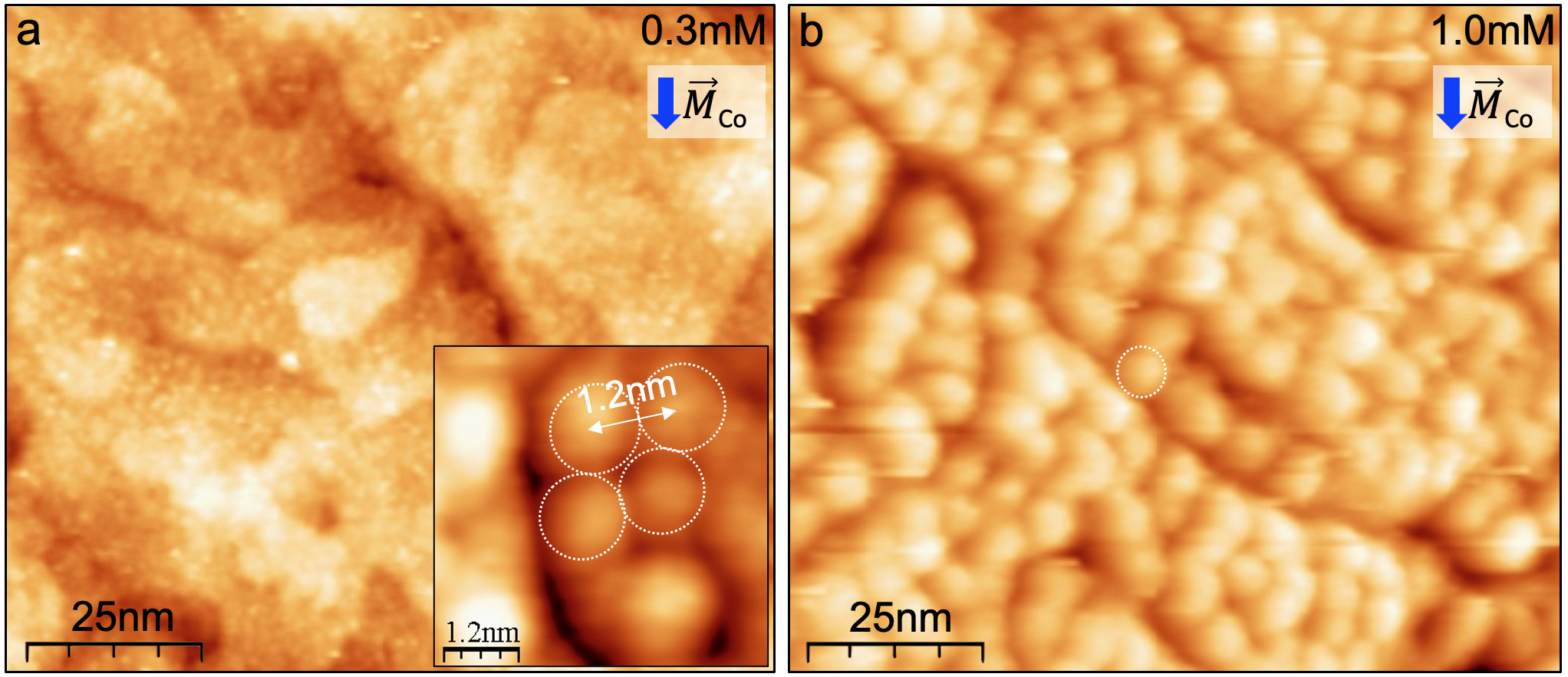}
	\caption{\label{FIG3}  a,b) STM images of L-PA on  Au/Co/Au: a) 0.3~mM. The inset reveals the characteristic intermolecular distance of 1.2~nm for self-assembled monolayer films \cite{HaAu2020}.  b) 1~mM  PA concentrations reveling formation of clusters on the surface. The tunneling conditions for both  images were V$_b$=+1.6~V, I$_t$=0.26~nA and $\vec{M}_{Co}=\downarrow$.}
\end{figure}

In order to obtain well self-assembled molecular monolayers, the concentration of PA  molecules should be considered in order to avoid clustering of molecules. The STM images in Fig.~\ref{FIG3} summarize this dependence on different PA concentrations.  
Fig.~\ref{FIG3}a) reveals    self-assembled monolayers (SAM)  of L-PA on terraces of Au/Co/Au(111) nanostructures, obtained  by drop-casting of the L-PA solution  with a concentration of 0.3~mM. The inset  indicates bright and hexagonally arranged protrusions with an average distance of 1.2~nm which is characteristic for a well-ordered SAM structure of PA molecules on Au(111) \cite{HaAu2020}. In our former work on Au(111) films with larger terraces but with using solutions of similar concentrations, we were able to resolve even submolecular structures, as shown in Fig.~\ref{FIG1}c) \cite{HaAu2020}. Therefore, also the roughness of the Au surface is crucial and triggers  the ordering process of the PA molecules on the Au(111) terraces.
 
Increasing the molecules concentration in the solution to 1~mM, a well-ordered cluster structure of L-PA was observed as shown in Fig.~\ref{FIG3}b) and introduced in the context of Fig.~\ref{FIG2}. 
From the cluster diameter and height distribution histograms, shown in Fig.~\ref{FIG2}c) and d), the clusters are almost spherical with a diameter of  6-7~nm, which is larger than the size of a single L-PA molecule. It was reported that, polymorphism (clustering) is easily triggered by the concentration of a solution \cite{Ha2011}.  Apparently, without any protection of the cysteine ligands in highly concentrated PA solutions, sulfur bonds are formed between the PA molecules resulting in clustering. Although such hedgehog-like spherical clusters (see inset in Fig.~\ref{FIG4}c) are still chiral, the molecules are differently oriented and coupled to the Au surface  after adsorption.

\subsection{Probing the spin-polarization with STS}
\begin{figure}[t]
	\includegraphics[width=\columnwidth]{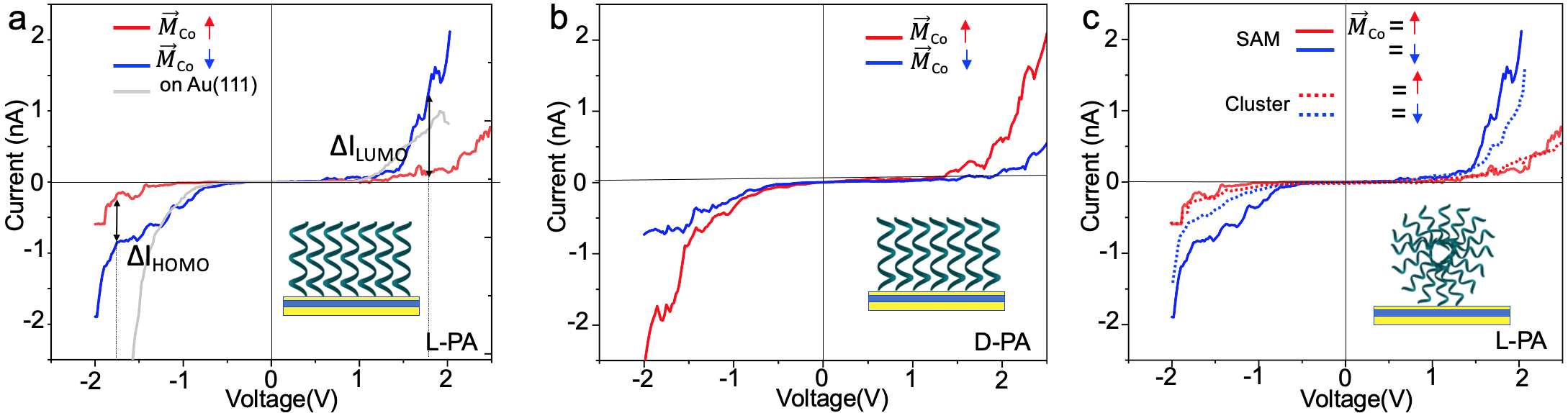}
	\caption{\label{FIG4}  I(V) curves of helical PA molecules adsorbed on magnetic Au/Co/Au substrates: a)  L-PA SAM (0.3~mM) on Au/Co/Au for both directions of magnetization. The I(V) curve for 36-mer L-PA on Au(111) grown on a mica substrate is shown in grey as a reference. The vertical lines denote the positions of the HOMO and LUMO state for PA \cite{HaAu2020}. b) D-PA SAM (0.3~mM) on Au/Co/Au for both directions of magnetization. c) L-PA SAM for 0.3~mM and clusters for 1~mM concentrations. All measurements were performed at 300~K with a setpoint of +1.6~V and 0.2~nA. The insets in a)-c) sketch the structure of SAMs (L-PA and D-PA) and a cluster of L-PA on the Au/Co/Ao nanostructure, respectively.}
\end{figure}

In order to confirm the chemical composition and presence of helicity of the PA molecules  and to study their electronic  transmission properties, we have performed locally I(V)-spectroscopy. All the curves displayed in Fig.~\ref{FIG4} for the Au/Co/Au/PA heterostructure show strong similarity to the I(V)-curves for highly ordered SAM of L-PA monolayers adsorbed on Au(111) films grown on mica proving a good quality of the heterostructure interfaces \cite{HaAu2020}.
The band gap for these 31-mer PA molecules is about 1.5~eV and very similar to the gap size measured for the 36-mer PA (displayed as a grey line Fig.~\ref{FIG4}a), indicating that  the PA molecules are chemically not altered. While the asymmetry of the I(V)-curves for PA on Au/Co/Au  is less pronounced in comparison to our findings for PA molecules on Au(111) films on mica and HOPG \cite{Ha2019,HaAu2020}, we have observed a strong dependence of the I(V) curves asymmetry on the Co magnetization direction.

In the  experiment shown in Fig.~\ref{FIG4}a) and b) we have adsorbed L-PA and D-PA molecules from 0.3~mM solutions on Au/Co/Au with different directions of the Co film magnetization. The line shape of the four I(V) curves is similar, which again confirms  similar adsorption geometries of PA molecules. However, the absolute currents are different. While in the case of L-PA the transmission is higher for the $\vec{M}_{Co}=\downarrow$ configuration, an opposite  behavior is found for D-PA for the same magnetization direction.  This finding directly demonstrates the spin-valve effect at the heterostructure interface of Au/Co/Au/PA, depending on the interplay between the chirality of the molecules and the magnetization direction of the Co-film.  E.g., for L-PA molecules, a higher transmission is found for both positive and negative bias voltage conditions for $\vec{M}_{Co}=\downarrow$, which is explained by a lower magnetoresistance $R_{MR}$ at the interface between  the metallic thin film system  and molecular layer. For the opposite direction of the Co-magnetization, the current is significantly lower. Taking into account that the strength of magnetization of the Co-film is the same for both orientations, the spin-polarization $P$ of the electrons within the heterostructure  can be derived from the relative difference of the tunneling current with respect to the direction of the magnetization ($\vec{M}_{Co}$=$\uparrow,\downarrow$), $ P= \frac{I_{\uparrow}-I_{\downarrow}}{I_{\uparrow}+I_{\downarrow}}$ \cite{Mondal2021} .

In   the case of L-PA, forming ordered SAMs, the spin polarization ratio of the itinerant electrons, averaged over the occupied and unoccupied states, was obtained to be  $P\approx 75~\%$ which is in reasonable agreement with results in previous studies \cite{Ghosh2020}.  For D-PA this value is reduced to 63~\% which  might by related to imperfections during the synthesis of D-cysteine and, subsequently, less-well ordered SAM structures of D-PA \cite{Mathew2014}. 
Besides, also the adsorption morphology  of the same helical molecules  can result in different spin-polarizations. As obvious from Fig.~\ref{FIG4}c), comparing L-PA SAMs with L-PA clusters, for the latter the spin-polarization was found to be only $P \approx 60~\%$, although  it was reported that a higher spin polarization is obtained by using longer and/or multilayers of PA molecules \cite{Haipeng2020}. The clusters in our case resemble locally a ''bilayer'' case  compared to a  SAM, however, the lateral arrangement of the molecules on the surface is different, which seems to be a decisive factor: the tunneling process always occurs via several molecules even in the SAM structure, e.g. a circular area with a diameter of 10~nm is probed, which is roughly the size of the clusters but with different spin polarizations. This finding indicates, that the CISS effect is rather a cooperative effect (triggered by several molecules), where the relative long axis molecule orientation of neighboring molecules is important. In the case of clusters, not all molecules are aligned perpendicularly to the surface (see inset in Fig.~\ref{FIG4}c). 
The interdigitation (rotational alignment)  found for PA on HOPG and Au(111) on mica indicates that rotational ordering seems to be  beneficial for obtaining high spin polarizations \cite{Ha2019}. The ordering of the PA molecules on the Au surface in our experiments  is inevitably related to the electronic coupling of PA to a nanostructure substrate  with a defined spin-injection into the Au-film from/into the Co layer below, which is relevant for  surface  magnetism induced by a proximity  of chiral molecules and the hybrid   spin-valve effect  \cite{BenDor2017}. This conclusion is in agreement with our finding of a charge-ordered state for PA molecules on Au(111) films grown on mica \cite{HaAu2020}.
Conversely, our results demonstrate also that the CISS effect is robust and appears also in less ordered (clustered) molecular structures. 

Moreover, closer analysis of the I(V)-curves shows, that the spin-polarization ratio deduced from the lowest unoccupied molecular orbitals (LUMOs) at positive bias voltages is significantly larger as compared to the spin-polarization derived from the transmission of the electrons along the highest occupied molecular orbitals (HOMO) ($\Delta I_{LUMO} \approx 2 \Delta I_{HOMO}$). This finding holds for both the ordered SAM structure and cluster phase and it seems  orbital overlap of LUMO is resulting in a higher electron mobility giving a higher polarization ratio.

The above mentioned calculation scheme of the spin polarization depends  on the magnetization orientation and also magnitude \cite{Mondal2021}. Therefore, the spin-polarization derived from the difference of the tunneling current through the PA molecules for both magnetization orientations provides an information about the spin polarization in the entire heterostructure rather than in the  helical molecules alone. For a more quantitative analysis, the MR effect across the interface needs to be considered, as we will show in the following paragraph.

\subsection{STM probing of the spin polarization by sequential switching of the magnetization}

\begin{figure}[t]
	\includegraphics[width=\columnwidth]{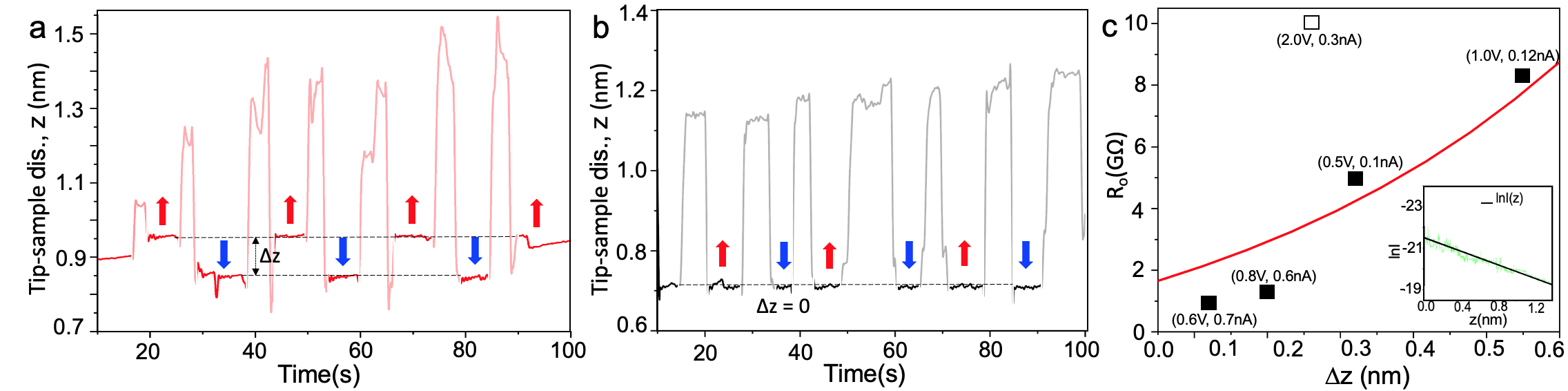}
	\caption{\label{FIG5} a) Tip-sample distance variation $\Delta z$ versus time during switching of the Co-magnetization for  L-PA SAM. The arrows denote the orientation of the Co-film magnetization, $\vec{M}_{Co}$, after the external magnet was moved away from the sample. b) Reference measurement. Tip-sample distance $z$ as a function of the Co-magnetization for a PA free Au/Co/Au substrate. c) The variation  of $\Delta$z-values for different set-point resistances $R_0$. The corresponding bias voltage and tunneling current are given in brackets ($V_b, I_t$). The red curve originates from a modeling using a simple tunneling model across the spin-valve heterojunction (see text). The data marked by $\square$ represents a set-point with a larger electric field across the tunneling junction. Inset:  Tunneling current I(z) versus tip-sample distance. Set point V$_b$=1.6~V, I$_t$=0.2~nA.}
\end{figure}

In Fig.~\ref{FIG5} we show the results of sequential magnetic field switching experiments, where the (non-spin-polarized) STM tip was kept in the tunneling regime at the same spot on the sample surface, while $\vec{M}_{Co}$ was sequentially switched up and down by means of an external magnet.  As pointed out above, due to the residual drift effects present during  the ambient STM measurements, the experiments were only possible in the constant current mode (closed-loop feedback).
Both the ability to grow homogeneous SAM molecular structures from a solution and to be able to switch the magnetization direction imposes in this case the  use of an ambient STM setup.  The average island diameter of around 20~nm limits the time and number of switching cycles of the Co thin film magnetization, which in our case  took  around 20 seconds for one back and forth switching event. 

In our experiment, the external magnet was  moved close to the sample for a few seconds (5-10~s). After retracting it, the magnet was oppositely oriented before the next approach. During this process, the STM remains in a constant current mode and electron transport occurs along the junction composed  of the entire hybrid structure consisting of  the helical molecules and the metallic substrate and the tunneling gap.  During the time, when the magnet was in proximity to the hybrid structure (near-field of the tunneling junction), strong amplitudes for the z-piezo were observed. After retracting  the magnet, the z-signal goes back to a  certain level. Most importantly, the final z-position of the STM piezo for these plateaus correlates with the Co film magnetization, indicated by the arrows, with $\Delta z \neq 0$ (see Fig.~\ref{FIG5}a). We have performed a similar magnetization switching sequence for the PA free Au/Co/Au system,  where $\Delta z =0$ (see Fig.~\ref{FIG5}b). This shows that the variation of the STM piezo is related to the spin polarization of the itinerant electrons across the  hybrid system consisting of the magnetic thin film system and the molecular layer at its surface. 

In order to evaluate the spin polarization from the $\Delta z$-signal, we assume that the resistance $R_0$ across the entire junction, according to the sketch in Fig.~\ref{FIG1}c), is composed of  the  gap tunneling resistance $R_{tunnel}$, the resistance of the molecules $R_{PA}$, the resistance of the Au/Co/Au substrate $R_{Au/Co/Au}$ and the contribution of the spin-dependent magnetoresistance across the hybrid structure $R_{MR}$.
For  typical set-points ($V_b, I_t$) used in our  two terminal transport experiments the total resistance is of the order of $R_0 \approx ~1~V/1~nA=1~G\Omega$. The resistance of the metallic nanostructure is in the order of $R_{Au/Co/Au}=10~\Omega$ and for the PA molecules we have obtained from mechanically break junction experiments $R_{PA}\approx 10~M\Omega$   \cite{Slawig2020}, thus these two smaller contributions originating from the magnetic substrate and PA molecules are negligible. 
Moreover, bias voltage and tunneling current were kept constant during the switching cycles of the magnetization, hence the resistance $R_0$ across the entire junction remains constant.  We assume therefore, that the change of the tunneling resistance (for a given STM set-point), induced by the different sample-tip distance $\Delta z$, is balanced  by the spin-valve effect: $\Delta R_{tunnel}(\Delta z)=\Delta R_{MR}$. 

A reasonable and justified assumption can be taken further that  during the change of the tip-sample distance z  the number of molecules involved in the tunneling process is not significantly changing and that the molecular orbital structure is not altered by varying the electric field across the junction. Moreover, the MR  should remain the same irrespective of the STM set-points, i.e. $R_{tunnel}$. Therefore, the absolute values of only $\Delta R_{tunnel}$  should be dependent on the STM set point, as shown in Fig.~\ref{FIG5}c). 

The average spin polarization $P$ can be deduced from the relative change of the resistance $\Delta R / R_{MR}= 2P^2/(1-P^2)$, where $\Delta R=\Delta R_{tunnel}$ is the difference of the resistance for the parallel and antiparallel magnetization orientation of the Co layer \cite{Tsymbal2003}. The reference resistance is given by $R_{MR}=R_0 - R_{tunnel}$. Keeping in mind that the tunneling resistance is approximated by $R_{tunnel}(z)=V_b I_t^{-1} e^{1.025 \sqrt{\Phi_{ap}}z}$, the resistance $R_0$ defined by the set point is related to $\Delta z$ as: $R_0=R_{tunnel}[\frac{(1-P^2)}{2P^2}(e^{1.025 \sqrt{\Phi_{ap}} \Delta z}-1)+1]$. The red solid line in Fig.~\ref{FIG5}c) shows a fit to various data points of $\Delta z$ for different STM set-points.  The trend seen in our experiment is reproduced assuming a spin polarization of $P \approx 40~\%$ in a reasonable agreement with the STS results. The apparent work function is $\Phi_{ap}=3.5~eV$ in consistency with the value obtained from fitting the I(z)-curve shown as inset in Fig.~\ref{FIG5}c).  Obviously, the $\Delta z$-change measured for large bias voltage values (open data point) are not covered by this simple theory at all.  As pointed out in the calculations presented in Ref.~\cite{Cristancho2010} and supported by recent transport measurements \cite{Slawig2020}, the molecular orbital structure seems to be significantly  affected by  high electric fields.

According to the model of Julliere \cite{Tsymbal2003}, the spin polarization $P$ can be written as the product of the spin-polarization of the Co-film, $P_{Co}$, and the spin-polarization of the molecules, $P_{PA}$. If we take spin polarization of 40\% to almost 60\%, reported for Co(0001) films and Co/Pt  multilayers   \cite{Getzlaff1996, Rajanikanth2010}, this model is giving a spin polarization in $P_{PA}$=$P$/$P_{Co}$ between 70\% and close to 100\%. Therefore, by knowing exactly the spin polarization of the magnetic field, the building-up of spin polarization in helical molecules is accessible.

\section{Conclusions}
The CISS effect was studied in detail by an ambient non-spin-polarited STM epitaxial magnetic  Au/Co/Au/PA hybrid structures as a function of the Co film magnetization direction as well as the helicity and coverage density of PA molecules. Contrarily to former studies \cite{Ziv2019,Ghosh2020} our high spatial and energy resolution of the STM setup allows the correlation of structural details with the electron transmission properties across the hybrid junction. Thereby, by analyzing the tunneling currents for both  constant current and constant height modes  for $\vec{M}_{Co}=\uparrow$ and $\vec{M}_{Co}=\downarrow$ configurations the spin-polarizations were quantified in detail.

The highest spin polarizations were found for well-ordered PA monolayer structures, where the cysteine termination of the PA molecule binds to the Au surface and the PA molecules form a self-assembled monolayer. Therefore, the current induced spin selectivity in helical molecules depends on the helical molecules ordering type proving that it is a cooperative effect and  also on the surface roughness of the capping layer in Au/Co/Au nanostructures. This is an important finding never reported before up to our knowledge. 

The differences of the tunneling currents for positive and negative bias voltage conditions were analyzed in such hybrid nanostructures. For both, the SAM and cluster types of PA-molecules ordering, the spin polarization obtained for electrons tunneling along the unoccupied states of the molecules is larger compared to electron transmissions along the highest occupied states.  We correlate this finding with a  larger orbital overlap resulting in a higher electron mobility along this molecular channel. 

The high sensitivity of the tunneling current in Au/Co/Au/PA hybrid nanostructure masured as a function of the tip-sample distance was used to quantify the magnetoresistance response. Thereby, the sequential switching of the Co-layer magnetization changes to the z-position of the STM tip over th sample surface.  Based on Julli\`{e}re's  model, the spin polarization was derived from tip-sample distance change   $\Delta z$. This opens a way  to determine  the spin polarization of buried magnetic layers or, if the spin-polarization of the magnetic layer is known, the  spin-polarization in helical molecules and maybe of all other chiral (non-helical) hybrid systems.  This hypothesis however needs further studies.

\vspace{1cm}


\begin{thebibliography}{9}

\bibitem[Xie \latin{et~al.}(2011)Xie, Markus, Cohen, Vager, Gutierrez, and
Naaman]{Xie2011}
Xie,~Z.; Markus,~T.~Z.; Cohen,~S.~R.; Vager,~Z.; Gutierrez,~R.; Naaman,~R.
{Spin Specific Electron Conduction through DNA Oligomers}. \emph{Nano
	Letters} \textbf{2011}, \emph{11}, 4652--4655\relax
\mciteBstWouldAddEndPuncttrue
\mciteSetBstMidEndSepPunct{\mcitedefaultmidpunct}
{\mcitedefaultendpunct}{\mcitedefaultseppunct}\relax
\EndOfBibitem
\bibitem[Naaman \latin{et~al.}(2019)Naaman, Paltiel, and Waldeck]{Naaman2019}
Naaman,~R.; Paltiel,~Y.; Waldeck,~D.~H. {Chiral molecules and the electron
	spin}. \emph{Nature Reviews Chemistry} \textbf{2019}, \emph{3},
250--260\relax
\mciteBstWouldAddEndPuncttrue
\mciteSetBstMidEndSepPunct{\mcitedefaultmidpunct}
{\mcitedefaultendpunct}{\mcitedefaultseppunct}\relax
\EndOfBibitem
\bibitem[Naaman \latin{et~al.}(2020)Naaman, Paltiel, and Waldeck]{Naaman2020}
Naaman,~R.; Paltiel,~Y.; Waldeck,~D.~H. {Chiral Molecules and the Spin
	Selectivity Effect}. \emph{The Journal of Physical Chemistry Letters}
\textbf{2020}, \emph{11}, 3660--3666\relax
\mciteBstWouldAddEndPuncttrue
\mciteSetBstMidEndSepPunct{\mcitedefaultmidpunct}
{\mcitedefaultendpunct}{\mcitedefaultseppunct}\relax
\EndOfBibitem
\bibitem[Waldeck \latin{et~al.}(2021)Waldeck, Naaman, and Paltiel]{Waldeck2021}
Waldeck,~D.~H.; Naaman,~R.; Paltiel,~Y. {The spin selectivity effect in chiral
	materials}. \emph{APL Materials} \textbf{2021}, \emph{9}, 040902\relax
\mciteBstWouldAddEndPuncttrue
\mciteSetBstMidEndSepPunct{\mcitedefaultmidpunct}
{\mcitedefaultendpunct}{\mcitedefaultseppunct}\relax
\EndOfBibitem
\bibitem[Sukenik \latin{et~al.}(2020)Sukenik, Tassinari, Yochelis, Millo,
Baczewski, and Paltiel]{Nir2020}
Sukenik,~N.; Tassinari,~F.; Yochelis,~S.; Millo,~O.; Baczewski,~L.~T.;
Paltiel,~Y. {Correlation between Ferromagnetic Layer Easy Axis and the Tilt
	Angle of Self Assembled Chiral Molecules}. \emph{Molecules} \textbf{2020},
\emph{25}\relax
\mciteBstWouldAddEndPuncttrue
\mciteSetBstMidEndSepPunct{\mcitedefaultmidpunct}
{\mcitedefaultendpunct}{\mcitedefaultseppunct}\relax
\EndOfBibitem
\bibitem[Meirzada \latin{et~al.}(2021)Meirzada, Sukenik, Haim, Yochelis,
Baczewski, Paltiel, and Bar-Gill]{Meirzada2021}
Meirzada,~I.; Sukenik,~N.; Haim,~G.; Yochelis,~S.; Baczewski,~L.~T.;
Paltiel,~Y.; Bar-Gill,~N. {Long-Time-Scale Magnetization Ordering Induced by
	an Adsorbed Chiral Monolayer on Ferromagnets}. \emph{ACS Nano} \textbf{2021},
\emph{15}, 5574--5579\relax
\mciteBstWouldAddEndPuncttrue
\mciteSetBstMidEndSepPunct{\mcitedefaultmidpunct}
{\mcitedefaultendpunct}{\mcitedefaultseppunct}\relax
\EndOfBibitem
\bibitem[Koyel \latin{et~al.}(2018)Koyel, Oren, Francesco, Eyal, Shira, Amir,
See-Hun, P., Soumyajit, Leeor, Tomasz, Ron, and Yossi]{Gosh2018}
Koyel,~B.-G.; Oren,~B.~D.; Francesco,~T.; Eyal,~C.; Shira,~Y.; Amir,~C.;
See-Hun,~Y.; P.,~P. S.~S.; Soumyajit,~S.; Leeor,~K.; Tomasz,~B.~L.; Ron,~N.;
Yossi,~P. {Separation of enantiomers by their enantiospecific interaction
	with achiral magnetic substrates}. \emph{Science} \textbf{2018}, \emph{360},
1331--1334\relax
\mciteBstWouldAddEndPuncttrue
\mciteSetBstMidEndSepPunct{\mcitedefaultmidpunct}
{\mcitedefaultendpunct}{\mcitedefaultseppunct}\relax
\EndOfBibitem
\bibitem[Tassinari \latin{et~al.}(2019)Tassinari, Steidel, Paltiel, Fontanesi,
Lahav, Paltiel, and Naaman]{Tassinari2019}
Tassinari,~F.; Steidel,~J.; Paltiel,~S.; Fontanesi,~C.; Lahav,~M.; Paltiel,~Y.;
Naaman,~R. {Enantioseparation by crystallization using magnetic substrates}.
\emph{Chemical Science} \textbf{2019}, \emph{10}, 5246--5250\relax
\mciteBstWouldAddEndPuncttrue
\mciteSetBstMidEndSepPunct{\mcitedefaultmidpunct}
{\mcitedefaultendpunct}{\mcitedefaultseppunct}\relax
\EndOfBibitem
\bibitem[Ghosh \latin{et~al.}(2020)Ghosh, Mishra, Avigad, Bloom, Baczewski,
Yochelis, Paltiel, Naaman, and Waldeck]{Ghosh2020}
Ghosh,~S.; Mishra,~S.; Avigad,~E.; Bloom,~B.~P.; Baczewski,~L.~T.;
Yochelis,~S.; Paltiel,~Y.; Naaman,~R.; Waldeck,~D.~H. {Effect of Chiral
	Molecules on the Electron's Spin Wavefunction at Interfaces}. \emph{The
	Journal of Physical Chemistry Letters} \textbf{2020}, \emph{11},
1550--1557\relax
\mciteBstWouldAddEndPuncttrue
\mciteSetBstMidEndSepPunct{\mcitedefaultmidpunct}
{\mcitedefaultendpunct}{\mcitedefaultseppunct}\relax
\EndOfBibitem
\bibitem[Liu \latin{et~al.}(2019)Liu, Wang, Wang, Shi, Gao, Feng, Deng, Hu,
Lochner, Schlottmann, von Moln'ar, Li, Zhao, of~Physics, University,
Tallahassee, Florida, Usa, of~Superlattices, Microstructures,
of~Semiconductors, of~Sciences, Beijing, China., for Condensed
Matter~Physics, of~Metal~Physics, of~Physical~Sciences, and
of~Technology]{Liu2019}
Liu,~T. \latin{et~al.}  {Spin selectivity through chiral polyalanine monolayers
	on semiconductors}. \emph{arXiv: Applied Physics} \textbf{2019}, \relax
\mciteBstWouldAddEndPunctfalse
\mciteSetBstMidEndSepPunct{\mcitedefaultmidpunct}
{}{\mcitedefaultseppunct}\relax
\EndOfBibitem
\bibitem[Naaman \latin{et~al.}(2019)Naaman, Waldeck, and Paltiel]{NaamanWY2019}
Naaman,~R.; Waldeck,~D.~H.; Paltiel,~Y. {Chiral molecules-ferromagnetic
	interfaces, an approach towards spin controlled interactions}. \emph{Applied
	Physics Letters} \textbf{2019}, \emph{115}, 133701\relax
\mciteBstWouldAddEndPuncttrue
\mciteSetBstMidEndSepPunct{\mcitedefaultmidpunct}
{\mcitedefaultendpunct}{\mcitedefaultseppunct}\relax
\EndOfBibitem
\bibitem[Nguyen \latin{et~al.}(2019)Nguyen, Solonenko, Selyshchev, Vogt, Zahn,
Yochelis, Paltiel, and Tegenkamp]{Ha2019}
Nguyen,~T. N.~H.; Solonenko,~D.; Selyshchev,~O.; Vogt,~P.; Zahn,~D. R.~T.;
Yochelis,~S.; Paltiel,~Y.; Tegenkamp,~C. {Helical Ordering of
	$\alpha$-l-Polyalanine Molecular Layers by Interdigitation}. \emph{The
	Journal of Physical Chemistry C} \textbf{2019}, \emph{123}, 612--617\relax
\mciteBstWouldAddEndPuncttrue
\mciteSetBstMidEndSepPunct{\mcitedefaultmidpunct}
{\mcitedefaultendpunct}{\mcitedefaultseppunct}\relax
\EndOfBibitem
\bibitem[Nguyen \latin{et~al.}(2020)Nguyen, Sharma, Slawig, Yochelis, Paltiel,
Zahn, Salvan, and Tegenkamp]{HaAu2020}
Nguyen,~T. N.~H.; Sharma,~A.; Slawig,~D.; Yochelis,~S.; Paltiel,~Y.; Zahn,~D.
R.~T.; Salvan,~G.; Tegenkamp,~C. {Charge-Ordered $\alpha$-Helical Polypeptide
	Monolayers on Au(111)}. \emph{The Journal of Physical Chemistry C}
\textbf{2020}, \emph{124}, 5734--5739\relax
\mciteBstWouldAddEndPuncttrue
\mciteSetBstMidEndSepPunct{\mcitedefaultmidpunct}
{\mcitedefaultendpunct}{\mcitedefaultseppunct}\relax
\EndOfBibitem
\bibitem[Nguyen \latin{et~al.}(2020)Nguyen, Xue, and Tegenkamp]{HaDL2020}
Nguyen,~T. N.~H.; Xue,~S.; Tegenkamp,~C. {Heterochiral Dimer Formation of
	$\alpha$-L- and $\alpha$-D-Polyalanine Molecules on Surfaces}. \emph{The
	Journal of Physical Chemistry C} \textbf{2020}, \emph{124},
11075--11080\relax
\mciteBstWouldAddEndPuncttrue
\mciteSetBstMidEndSepPunct{\mcitedefaultmidpunct}
{\mcitedefaultendpunct}{\mcitedefaultseppunct}\relax
\EndOfBibitem
\bibitem[Aragon{\`e}s \latin{et~al.}(2017)Aragon{\`e}s, Medina, Ferrer-Huerta,
Gimeno, Teixid{\'o}, Palma, Tao, Ugalde, Giralt, D{\'\i}ez-P{\'e}rez, and
Mujica]{Aragones2017}
Aragon{\`e}s,~A.~C.; Medina,~E.; Ferrer-Huerta,~M.; Gimeno,~N.;
Teixid{\'o},~M.; Palma,~J.~L.; Tao,~N.; Ugalde,~J.~M.; Giralt,~E.;
D{\'\i}ez-P{\'e}rez,~I.; Mujica,~V. {Measuring the Spin-Polarization Power of
	a Single Chiral Molecule}. \emph{Small} \textbf{2017}, \emph{13},
1602519\relax
\mciteBstWouldAddEndPuncttrue
\mciteSetBstMidEndSepPunct{\mcitedefaultmidpunct}
{\mcitedefaultendpunct}{\mcitedefaultseppunct}\relax
\EndOfBibitem
\bibitem[Slawig \latin{et~al.}(2020)Slawig, Nguyen, Yochelis, Paltiel, and
Tegenkamp]{Slawig2020}
Slawig,~D.; Nguyen,~T. N.~H.; Yochelis,~S.; Paltiel,~Y.; Tegenkamp,~C.
{Electronic transport through single polyalanine molecules}. \emph{Physical
	Review B} \textbf{2020}, \emph{102}, 115425--\relax
\mciteBstWouldAddEndPuncttrue
\mciteSetBstMidEndSepPunct{\mcitedefaultmidpunct}
{\mcitedefaultendpunct}{\mcitedefaultseppunct}\relax
\EndOfBibitem
\bibitem[Kiran \latin{et~al.}(2016)Kiran, Cohen, and Naaman]{Kiran2016}
Kiran,~V.; Cohen,~S.~R.; Naaman,~R. {Structure dependent spin selectivity in
	electron transport through oligopeptides}. \emph{The Journal of Chemical
	Physics} \textbf{2016}, \emph{146}, 092302\relax
\mciteBstWouldAddEndPuncttrue
\mciteSetBstMidEndSepPunct{\mcitedefaultmidpunct}
{\mcitedefaultendpunct}{\mcitedefaultseppunct}\relax
\EndOfBibitem
\bibitem[Ziv \latin{et~al.}(2019)Ziv, Saha, Alpern, Sukenik, Baczewski,
Yochelis, Reches, and Paltiel]{Ziv2019}
Ziv,~A.; Saha,~A.; Alpern,~H.; Sukenik,~N.; Baczewski,~L.~T.; Yochelis,~S.;
Reches,~M.; Paltiel,~Y. {AFM-Based Spin-Exchange Microscopy Using Chiral
	Molecules}. \emph{Advanced Materials} \textbf{2019}, \emph{31}, 1904206\relax
\mciteBstWouldAddEndPuncttrue
\mciteSetBstMidEndSepPunct{\mcitedefaultmidpunct}
{\mcitedefaultendpunct}{\mcitedefaultseppunct}\relax
\EndOfBibitem
\bibitem[Wortmann \latin{et~al.}(2001)Wortmann, Heinze, Kurz, Bihlmayer, and
Bl{\"u}gel]{Wortmann2001}
Wortmann,~D.; Heinze,~S.; Kurz,~P.; Bihlmayer,~G.; Bl{\"u}gel,~S. {Resolving
	Complex Atomic-Scale Spin Structures by Spin-Polarized Scanning Tunneling
	Microscopy}. \emph{Physical Review Letters} \textbf{2001}, \emph{86},
4132--4135\relax
\mciteBstWouldAddEndPuncttrue
\mciteSetBstMidEndSepPunct{\mcitedefaultmidpunct}
{\mcitedefaultendpunct}{\mcitedefaultseppunct}\relax
\EndOfBibitem
\bibitem[Ben~Dor \latin{et~al.}(2017)Ben~Dor, Yochelis, Radko, Vankayala,
Capua, Capua, Yang, Baczewski, Parkin, Naaman, and Paltiel]{BenDor2017}
Ben~Dor,~O.; Yochelis,~S.; Radko,~A.; Vankayala,~K.; Capua,~E.; Capua,~A.;
Yang,~S.-H.; Baczewski,~L.~T.; Parkin,~S. S.~P.; Naaman,~R.; Paltiel,~Y.
{Magnetization switching in ferromagnets by adsorbed chiral molecules without
	current or external magnetic field}. \emph{Nature Communications}
\textbf{2017}, \emph{8}, 14567\relax
\mciteBstWouldAddEndPuncttrue
\mciteSetBstMidEndSepPunct{\mcitedefaultmidpunct}
{\mcitedefaultendpunct}{\mcitedefaultseppunct}\relax
\EndOfBibitem
\bibitem[Carmeli \latin{et~al.}(2002)Carmeli, Skakalova, Naaman, and
Vager]{Carmeli2002}
Carmeli,~I.; Skakalova,~V.; Naaman,~R.; Vager,~Z. {Magnetization of Chiral
	Monolayers of Polypeptide: A Possible Source of Magnetism in Some Biological
	Membranes}. \emph{Angewandte Chemie International Edition} \textbf{2002},
\emph{41}, 761--764\relax
\mciteBstWouldAddEndPuncttrue
\mciteSetBstMidEndSepPunct{\mcitedefaultmidpunct}
{\mcitedefaultendpunct}{\mcitedefaultseppunct}\relax
\EndOfBibitem
\bibitem[Mondal \latin{et~al.}(2021)Mondal, Preuss, Sleczkowski, Das, Vantomme,
Meijer, and Naaman]{Mondal2021}
Mondal,~A.~K.; Preuss,~M.~D.; Sleczkowski,~M.~L.; Das,~T.~K.; Vantomme,~G.;
Meijer,~E.~W.; Naaman,~R. {Spin Filtering in Supramolecular Polymers
	Assembled from Achiral Monomers Mediated by Chiral Solvents}. \emph{J. Am.
	Chem. Soc.} \textbf{2021}, \emph{143}, 7189\relax
\mciteBstWouldAddEndPuncttrue
\mciteSetBstMidEndSepPunct{\mcitedefaultmidpunct}
{\mcitedefaultendpunct}{\mcitedefaultseppunct}\relax
\EndOfBibitem
\bibitem[Erskine and Stern(1973)Erskine, and Stern]{Erskine1973}
Erskine,~J.~L.; Stern,~E.~A. {Magneto-optic Kerr Effect in Ni, Co, and Fe}.
\emph{Phys. Rev. Lett.} \textbf{1973}, \emph{30}, 1329--1332\relax
\mciteBstWouldAddEndPuncttrue
\mciteSetBstMidEndSepPunct{\mcitedefaultmidpunct}
{\mcitedefaultendpunct}{\mcitedefaultseppunct}\relax
\EndOfBibitem
\bibitem[Sharma \latin{et~al.}(2020)Sharma, Matthes, Soldatov, Arekapudi,
B\"o~hm, Lindner, Selyshchev, Thi Ngoc~Ha, Mehring, Tegenkamp, Schulz, Zahn,
Paltiel, Hellwig, and Salvan]{Sharma2020}
Sharma,~A.; Matthes,~P.; Soldatov,~I.; Arekapudi,~S. S. P.~K.; B\"o~hm,~B.;
Lindner,~M.; Selyshchev,~O.; Thi Ngoc~Ha,~N.; Mehring,~M.; Tegenkamp,~C.;
Schulz,~S.~E.; Zahn,~D. R.~T.; Paltiel,~Y.; Hellwig,~O.; Salvan,~G. {Control
	of magneto-optical properties of cobalt-layers by adsorption of
	$\alpha$-helical polyalanine self-assembled monolayers}. \emph{J. Mater.
	Chem. C} \textbf{2020}, \emph{8}, 11822--11829\relax
\mciteBstWouldAddEndPuncttrue
\mciteSetBstMidEndSepPunct{\mcitedefaultmidpunct}
{\mcitedefaultendpunct}{\mcitedefaultseppunct}\relax
\EndOfBibitem
\bibitem[Thi Ngoc~Ha \latin{et~al.}(2011)Thi Ngoc~Ha, Gopakumar, and
Hietschold]{Ha2011}
Thi Ngoc~Ha,~N.; Gopakumar,~T.~G.; Hietschold,~M. {Polymorphism Driven by
	Concentration at the Solid--Liquid Interface}. \emph{The Journal of Physical
	Chemistry C} \textbf{2011}, \emph{115}, 21743--21749\relax
\mciteBstWouldAddEndPuncttrue
\mciteSetBstMidEndSepPunct{\mcitedefaultmidpunct}
{\mcitedefaultendpunct}{\mcitedefaultseppunct}\relax
\EndOfBibitem
\bibitem[Mathew \latin{et~al.}(2014)Mathew, Mondal, Moshe, Mastai, and
Naaman]{Mathew2014}
Mathew,~S.~P.; Mondal,~P.~C.; Moshe,~H.; Mastai,~Y.; Naaman,~R. {Non-magnetic
	organic/inorganic spin injector at room temperature}. \emph{Applied Physics
	Letters} \textbf{2014}, \emph{105}, 242408\relax
\mciteBstWouldAddEndPuncttrue
\mciteSetBstMidEndSepPunct{\mcitedefaultmidpunct}
{\mcitedefaultendpunct}{\mcitedefaultseppunct}\relax
\EndOfBibitem
\bibitem[Haipeng \latin{et~al.}()Haipeng, Jingying, Chuanxiao, Xin, Xihan,
Roman, J., Kai, C., and Valy]{Haipeng2020}
Haipeng,~L.; Jingying,~W.; Chuanxiao,~X.; Xin,~P.; Xihan,~C.; Roman,~B.;
J.,~B.~J.; Kai,~Z.; C.,~B.~M.; Valy,~V.~Z. {Spin-dependent charge transport
	through 2D chiral hybrid lead-iodide perovskites}. \emph{Science Advances}
\emph{5}, eaay0571\relax
\mciteBstWouldAddEndPuncttrue
\mciteSetBstMidEndSepPunct{\mcitedefaultmidpunct}
{\mcitedefaultendpunct}{\mcitedefaultseppunct}\relax
\EndOfBibitem
\bibitem[Tsymbal \latin{et~al.}(2003)Tsymbal, Mryasov, and
LeClair]{Tsymbal2003}
Tsymbal,~E.~Y.; Mryasov,~O.~N.; LeClair,~P.~R. {Spin-dependent tunnelling in
	magnetic tunnel junctions}. \emph{Journal of Physics: Condensed Matter}
\textbf{2003}, \emph{15}, R109--R142\relax
\mciteBstWouldAddEndPuncttrue
\mciteSetBstMidEndSepPunct{\mcitedefaultmidpunct}
{\mcitedefaultendpunct}{\mcitedefaultseppunct}\relax
\EndOfBibitem
\bibitem[Cristancho and Seminario(2010)Cristancho, and
Seminario]{Cristancho2010}
Cristancho,~D.; Seminario,~J.~M. {Polypeptides in alpha-helix conformation
	perform as diodes}. \emph{J. Chem. Phys.} \textbf{2010}, \emph{132},
065102\relax
\mciteBstWouldAddEndPuncttrue
\mciteSetBstMidEndSepPunct{\mcitedefaultmidpunct}
{\mcitedefaultendpunct}{\mcitedefaultseppunct}\relax
\EndOfBibitem
\bibitem[Getzlaff \latin{et~al.}(1996)Getzlaff, Bansmann, Braun, and
Schönhense]{Getzlaff1996}
Getzlaff,~M.; Bansmann,~J.; Braun,~J.; Schönhense,~G. Spin resolved
photoemission study of Co(0001) films. \emph{Journal of Magnetism and
	Magnetic Materials} \textbf{1996}, \emph{161}, 70--88\relax
\mciteBstWouldAddEndPuncttrue
\mciteSetBstMidEndSepPunct{\mcitedefaultmidpunct}
{\mcitedefaultendpunct}{\mcitedefaultseppunct}\relax
\EndOfBibitem
\bibitem[Rajanikanth \latin{et~al.}(2010)Rajanikanth, Kasai, Ohshima, and
Hono]{Rajanikanth2010}
Rajanikanth,~A.; Kasai,~S.; Ohshima,~N.; Hono,~K. Spin polarization of currents
in Co/Pt multilayer and Co-Pt alloy thin films. \emph{Appl. Phys. Lett.}
\textbf{2010}, \emph{97}, 022505\relax
\mciteBstWouldAddEndPuncttrue
\mciteSetBstMidEndSepPunct{\mcitedefaultmidpunct}
{\mcitedefaultendpunct}{\mcitedefaultseppunct}\relax
\end{thebibliography}

\end{document}